\documentstyle[preprint,aps]{revtex}

\begin{document}


\title{One and two dimensional tunnel junction arrays in weak Coulomb blockade regime - absolute accuracy in 
thermometry} 

\author{J. P. Pekola, L. J. Taskinen, and Sh. Farhangfar} 

\address{Department of Physics, University of Jyv{\"a}skyl{\"a}, P.O. Box 35 (Y5), FIN-40351 
Jyv{\"a}skyl{\"a}, Finland}

\date{\today}
\maketitle

\begin{abstract}
We have investigated one and two dimensional (1D and 2D) arrays of tunnel junctions in partial Coulomb 
blockade regime. The absolute accuracy of the Coulomb blockade thermometer is influenced by the external 
impedance of the array, which is not the same in the different topologies of 1D and 2D arrays. We demonstrate, 
both by experiment and by theoretical calculations in simple geometries, that the 1D structures are better in this 
respect. Yet in both 1D and 2D, the influence of the environment can be made arbitrarily small by making the 
array sufficiently large. 
\end{abstract}
\vspace{2.5 cm}


\newpage

Coulomb blockade thermometry (CBT) was invented five years ago \cite{jp94} and has since been established 
as accurate and practical means to determine absolute temperature \cite{cbts}. Until recently only one 
dimensional (1D) arrays were discussed. An interesting suggestion to use two dimensional (2D) arrays was put 
forward by Bergsten {\it et al.} in \cite{pd99}, where limitations in measurement rate and tolerance to 
fabrication failures were discussed and shown to be even less restrictive than in 1D arrays. A large 2D array 
consisting of 256 $\times$ 256 tunnel junctions yielded absolute accuracy of better than 0.3 \% at temperatures 
from two to four kelvin. This is somewhat better, although statistical variations from sample to sample in 2D arrays 
have not been reported, than what has been achieved with 1D arrays consisting of only 20 junctions in series (0.5 
\% standard deviation in absolute accuracy). Besides considering practical thermometry with sensors having not 
excessively many junctions, it is very interesting to understand the influence of the electromagnetic environment 
on tunnelling in arrays. In the case of a small tunnel junction and weak tunnelling ($R_{\rm T} \gg R_{\rm Q}$, 
with $R_{\rm T}$ the junction resistance and $R_{\rm Q} = h/4e^2 \simeq 6.5$ k$\Omega$) this has been 
treated by the phase correlation theory with the harmonic oscillator bath as the environment \cite{dev,ing}, and it 
has been recently extended to describe in detail double junction structures \cite{est2,far} and long 1D arrays 
\cite{far}. A general approach which includes strong tunnelling in 1D arrays is due to Golubev and Zaikin 
\cite{gz} with emphasis on the low temperature limit. Our purpose here is to show experimental data on a few 
topologically different sets of arrays, and to discuss the theory in some of the less complicated cases.

CBT is based on partial blockade of tunnelling in the regime where the charging energy of single electrons, 
$E_{\rm C}\equiv 2\frac{N-1}{N}\frac{e^2}{2C}$ ($N$ is the number of junctions  
and $C\equiv{ C_i}$ is capacitance of the {\it i}th junction in a {\it homogeneous} array), the thermal energy, 
$k_{\rm B}T$, and the electrostatic energy difference across the array, $eV$, where $V$ is the bias voltage, 
compete. The significant property is that the full conductance curve, $G/G_{\rm T}$, against $V$, or in more 
general terms against $v \equiv eV/Nk_{\rm B}T$,  can be calculated with a universal result for not too large 
values of the ratio $u\equiv E_{\rm C}/k_{\rm B} T$, which is the expansion parameter. Here, $G_{\rm T} 
\equiv R_{\rm T}^{-1}$ is the conductance of the array at large transport voltages. The basic result is the linear 
one, 
\begin{equation} \label{1}
G/G_{\rm T} = 1 - u g(v)
\end{equation}
with known universal corrections for not small values of $u$. Here $g(x) = [x \sinh (x) - 4 \sinh ^2 (x/2)]/8 \sinh^4 
(x/2)$ is the function introduced in Ref. \cite{jp94}. The main result is that the full width at half minimum in this 
linear regime, $V_{1/2,0}$, has the value
\begin{equation} \label{2}
V_{1/2,0}=5.439Nk_{\rm B}T/e
\end{equation}
with again a known correction $\Delta V_{1/2} = V_{1/2} - V_{1/2,0}$, which is proportional to the normalised 
depth of the conductance dip, $\Delta G / G_{\rm T}$: 
\begin{equation} \label{3}
\Delta V_{1/2}/V_{1/2,0} = 0.39211 \Delta G/G_{\rm T}.
\end{equation}
An important feature of CBT is that it is not significantly influenced by random background charges when $u \le 
1$.
The half width $V_{1/2}$, which is proportional to $T$ at small values of $u$, serves as an absolute measure of 
temperature. $\Delta G / G_{\rm T}$, in turn, is a secondary thermometric parameter, inversely proportional to 
$T$ (again in the linear regime):
\begin{equation} \label{4}
\Delta G / G_{\rm T} = E_{\rm C} /6k_{\rm B} T.
\end{equation}
The formulae (\ref{1} - \ref{4}) are obeyed quantitatively with high precision in 1D arrays consisting of a large 
number of junctions. Yet in simpler structures, with, e.g., $N < 10$ junctions in a 1D array, the influence of the 
external impedance on tunnelling is significant \cite{cbts,est2,far}, and the conductance dip broadens. The 
origin of this is that Eqs. (\ref{1} - \ref{4}) are based on the assumption that the arrays are perfectly voltage 
biased at the ends, which is not the case due to the nonzero external on-chip impedance. If we neglect the 
influence of the environment and non-sequential tunnelling, we can, however, analyse our arrays using Eqs. 
(\ref{1} - \ref{4}) with no extra corrections, if we assume that the structures are fairly uniform. In simple cases, 
like with $N$ junction 1D arrays, these results are easy to obtain analytically \cite{jp94,cbts}, but with 2D arrays 
we have had to resort to Monte Carlo  simulations.

Topologically different 1D and 2D arrays which we will discuss here are shown schematically in Fig 1. The bias is 
connected between the left and the right end (busbars). The practical $N$ junction 1D arrays used for Coulomb 
blockade thermometry consist of $M$ nominally identical arrays in parallel. The parallel connection does not 
affect the results of Eqs. (\ref{1} - \ref{4}), but it lowers the impedance of the sensors by $1/M$ and this way 
makes them more suitable for practical measurements \cite{cbts}. In the basic "aligned" 2D structure the 
"equipotential" islands of the neighbouring chains are connected to each other by a tunnel junction whose 
parameters, i.e., $R_{\rm T}$ and capacitance $C$, are equal to those of the rest of the junctions in the 
structure. In the "diagonal" 2D structure we denote the number of connections at each busbar by $M'$.

As the simplest illustrative example of comparing 1D and 2D structures, let us take 1D and 2D arrays with 
$N=2$, $M=2$. They are schematically shown in Fig. 1 (d). We denote such a 1D structure by {\bf II} and the 2D 
structure by {\bf H}. They were fabricated on nitridised silicon chips by standard electron beam lithography and 
two angle shadow evaporation of aluminum, with oxidation to create the tunnel barrier in between. The size of the 
junctions was nominally 0.2 $\times$ 0.6 $\mu$m$^2$, and three different geometries of both types of samples 
were employed to check the influence of the internal structure of the array and of its termination. For example, 
island size was varied from 1 $\mu$m up to 10 $\mu$m.

Four {\bf II} structures and seven {\bf H} structures were measured at $T = $ 4.25 K. The inset of Fig. 2 shows 
a typical example of the conductance of a {\bf II} structure and an {\bf H} structure fabricated simultaneously on 
the same chip. In both cases $\Delta G / G_{\rm T} \simeq 0.01$, which implies very small corrections from the 
linear result of Eqs. (\ref{1}) and (\ref{2}) (see Eq. (\ref{3})). Although not very apparent from the two curves 
in the inset of Fig. 2, the generic features are: (i) the dip of the {\bf H} structure is wider than that of the {\bf II} 
structure, (ii) the shapes of the dips are not the same; simple scaling of the two curves does not make them 
overlap, and (iii) the dips are nearly equally high, when the junction size is the same in {\bf II} and {\bf H}. As will 
be discussed in what follows, (i) - (iii) are not consistent with the simple sequential tunnelling result with zero 
external impedance.  The simple two junction result \cite{jp94} with no external impedance would predict 
$V_{1/2,0} =$ 4.0 mV at 4.25 K. The main result here, instead, is that $V_{1/2}$ of the {\bf II} and {\bf H} 
structures cluster around different values, both higher than $V_{1/2,0}$, i.e., at $4.53 \pm 0.02$ mV 
($V_{1/2}/V_{1/2,0} = 1.13 \pm 0.005$) for the {\bf II} and at $4.79 \pm 0.05$ mV ($V_{1/2}/V_{1/2,0} = 
1.20 \pm 0.01$) for {\bf H}, respectively. The internal geometry of the array did not influence the results. This 
result already suggests that the corrections to the basic results of Eqs. (\ref{1} - \ref{4}) are larger for 2D type 
structures.

And, to make a comparison between all different array types of Fig. 1, several samples of different types with 
$N=8$ were measured. The number of parallel connections was $M=9$ in 1D and in "aligned" 2D; the 
"diagonal" 2D arrays had $M'=4$. Simple theory (Eq. (\ref{2})) predicts $V_{1/2,0} =$ 16.0 mV. Typical 
conductance curves measured at 4.25 K are shown in Fig. 2 for 1D and "aligned" 2D arrays. Also for these 
samples $\Delta G / G_{\rm T} \sim 0.01$, being well in the linear regime of Eqs. (\ref{1}) and (\ref{2}).  The 
histogram of $V_{1/2}$ of all the samples measured is shown in Fig 3 (a). 
The 1D arrays show a $+4$ \% correction to $V_{1/2,0}$ ($V_{1/2} = 16.68 \pm 0.04$ mV on the average). 
The "diagonal" 2D arrays have a width of $V_{1/2} =18.20 \pm 0.14$ mV, i.e., it is 14 \% wider than the theory 
would predict. The "aligned" 2D arrays show $V_{1/2} = 19.0 \pm 0.27$ mV, which exceeds $V_{1/2,0}$ by 
19 \%. Thus, at least in the case of arrays with a relatively small charging peak ($\Delta G / G_{\rm T} \sim 1$ 
\%), the 1D structures are more suitable for thermometry if judged based on the faster decrease of the size ($N$) 
dependent correction observed. Few arrays with smaller junctions
were also measured: the difference from the simple theory becomes smaller for all arrays, but we have not done 
this systematically enough for quantitative conclusions. Our basic conclusion on $N$ dependence is further 
supported by the measurements of $N=20$, $M=21$ and $N=30$, $M=31$ 1D and "aligned" 2D arrays: the 
data are collected in Fig. 3.
 
To interpret the results we apply the existing phase correlation theory of single electron tunnelling 
\cite{ing,est2,far}. Before doing that we first see the relation of the conductance in structures {\bf H} and {\bf II} 
if the external impedance of these structures is assumed to be vanishingly small. We can then use the "orthodox" 
theory of single electron tunnelling \cite{avelik}, and derive the conductance curve either analytically in the high 
temperature limit ($u < 1$) \cite{jp94}, or numerically by a Monte Carlo simulation \cite{bakh,hirvi}. The 
obvious result is, that in both structures the conductance curve has the form given by Eq. (\ref{1}), but the 
capacitance seen by one of the islands to the bias lines is $2C$ in {\bf II}, whereas in {\bf H} it is $8C/3$. 
Thus, $E_{\rm C}$ of {\bf H} is smaller by a factor 3/4, and so is the depth $\Delta G / G_{\rm T} = u/6$. The 
width of the dip is not different in the two structures {\bf H} and {\bf II}. The corrections in $V_{1/2}$ observed 
experimentally are thus due to less trivial reasons.

The most obvious corrections to the basic orthodox result arise either from the higher order tunnelling events (co-
tunnelling and other non-sequential processes) or from the impedance of the environment. The first possibility can 
be ruled out easily: the higher order events are important in the limit of small tunnel resistances ($R_{\rm T} < 
R_{\rm Q} \simeq 6.5 $ k$\Omega$), but our results did not show correlation with the tunnel resistance of the 
samples, which varied from 15 k$\Omega$ up to 50 k$\Omega$. Therefore, we analysed our observations 
based on the phase correlation theory (i.e., perturbatively) in the {\bf H} and {\bf II} circuits. 

The tunnelling rates $\Gamma_j^{\pm}$ through the junction $j$ in the two directions ($\pm$) are given by 
\begin{equation} \label{5}
\Gamma_j^{\pm}(\delta F_j^{\pm}) = (1/e^2R_{\rm T}) \int_{-\infty}^{+\infty} dE {\frac{E}{1-\exp (-
E/k_{\rm B}T)}} P_j(-\delta F_j^{\pm} - E).
\end{equation} 
Here $\delta F_j^{\pm}$ is the free energy change in the tunnelling event and 
$P_j(E)=\frac{1}{2\pi \hbar}  \int_{-\infty}^{+\infty}dt e^{[ J_j(t) + i\frac{E}{\hbar}t ]}$ is the probability 
density of the tunnelling electron to exchange energy $E$ with its electromagnetic environment, 
where in turn, $J_j(t)$ is the phase-phase correlation function:
\begin{equation} \label{6}
J_j(t) = 2 \int_{0}^{+\infty} {\frac{d \omega} {\omega}} {\frac {{\rm Re}[Z_{{\rm t},j} 
(\omega)]}{R_{\rm K}}}\large \{ \coth ({\frac {\hbar \omega} {2k_{\rm B} T}})[\cos (\omega t) -1] - i 
\sin(\omega t)\large \}.
\end{equation}
The essential parameter is the real part of the impedance $Z_{{\rm t},j}$ seen by junction $j$. This can be 
obtained as the equivalent impedance of the surrounding circuit in parallel with the capacitance of this junction 
\cite{ing}. All other junctions are described as pure capacitors, based on the assumption of sequential 
tunnelling. 

To describe the environment of the junction circuit we have used the simple resistive impedance. The circuits of 
{\bf II} are identical to what has been described in Refs. \cite{est2,far} with resistance $R_{\rm e}$ at both 
ends of the double junction(s), see inset in Fig. 3 (c). Similarly we assume that there is impedance $R_{\rm e}$ 
connected at each termination of the {\bf H} structure, as shown also in the inset. The impedances ${\rm 
Re}[Z_{{\rm t},j} (\omega)]$ in Eq. (\ref{6}) can then be easily calculated and the result is
${\rm Re}[Z_{{\rm t},j}(\omega)] = {\frac {R_{\rm e}}{2}} {\frac{1}{1 + (\omega / \omega_{\rm 
c})^2}}$
for all the four junctions in {\bf II}, and
${\rm Re}[Z_{{\rm t},j}(\omega)] = {\frac {9R_{\rm e}}{16}} {\frac{1 + (1/3)(\omega / \omega_{\rm 
c})^2}{1 + (5/4)(\omega / \omega_{\rm c})^2 + (1/4)(\omega / \omega_{\rm c})^4}}$
for the symmetrically positioned four junctions in {\bf H}, and
${\rm Re}[Z_{{\rm t},j}(\omega)] = {\frac {R_{\rm e}}{4}} {\frac{1}{1 + (1/4)(\omega / \omega_{\rm 
c})^2}}$ for the interconnecting central junction in {\bf H}.  The cut-off frequency is defined as $\omega_{\rm 
c} \equiv (R_{\rm e}C)^{-1}$. We calculated the current voltage characteristics of the {\bf II} and {\bf H} 
structures as functions of $R_{\rm e}$ with two values, $C= 6$ fF and $C = 3$ fF, corresponding to the 
estimated range of capacitances of the samples measured, and at $T= 4.25$ K. Despite the fact that we used 
just resistive environment we can draw some, at least qualitative conclusions based on the results depicted in 
Fig. 3 (c), where the half width of the corresponding conductance dip from the calculation has been plotted 
against $R_{\bf e}$. Since the "unintentional" on-chip impedance is typically a fraction of the free space 
impedance $\simeq$ 377 $\Omega$, i.e., on the order of 100 $\Omega$, we can assume that we are close to or 
below the maximum of $V_{1/2}$ on the $R_{\rm e}$ scale for both {\bf II} and {\bf H} structures. Therefore 
the width of the {\bf H} structure well exceeds that of the {\bf II} structure. If we assume that we are exactly at the 
maximum in both cases, {\bf II} and {\bf H}, even the quantitative agreement between the measured and the 
calculated half widths is good: from the calculation we obtain $V_{1/2} = $ 4.9 mV and $V_{1/2} = $ 4.6 mV 
for {\bf H} and {\bf II} structures, respectively. The coincidence between experiment and theory may be partly 
accidental, but gives a ground for the observed difference in the measured characteristics of {\bf H} and {\bf II}. 
The additional qualitative observation that the influence of the environment is most pronounced for very weak 
Coulomb blockade ($\Delta G / G_{\rm T} \sim  0.01$) can be understood by noting that the position of the 
maximum of $V_{1/2}$ on the $R_{\rm e}$ scale is approximately inversely proportional to $E_{\rm C}$, but 
its magnitude is almost unchanged. The analysis of the larger 2D arrays in a dissipative environment is markedly 
more complicated and we have not tried that. The larger scatter in the data of 2D arrays in Fig. 3 (b) may indicate 
stronger $R_{\rm e}$ dependence than in 1D, similarly to what is predicted in Fig. 3 (c) for $N=2$. 

In summary, we have investigated the influence of the cross connections in arrays of small tunnel junctions on their 
current voltage characteristics. We have observed that the multiply linked circuit structure in 2D arrays leads to 
strong influence of the external electromagnetic environment on the tunnelling junction imbedded in the array. This 
affects drastically the thermometric properties of not very extended junction arrays in Coulomb blockade 
thermometry in favour of 1D structures. 

We thank Antti Manninen, Andrei Zaikin, Per Delsing, Jari Kinaret, Klavs Hansen and Tobias Bergsten for 
discussions, and the National Graduate School in Materials Physics for support.

\newpage
{FIGURE CAPTIONS}
{\vskip 0.5cm}

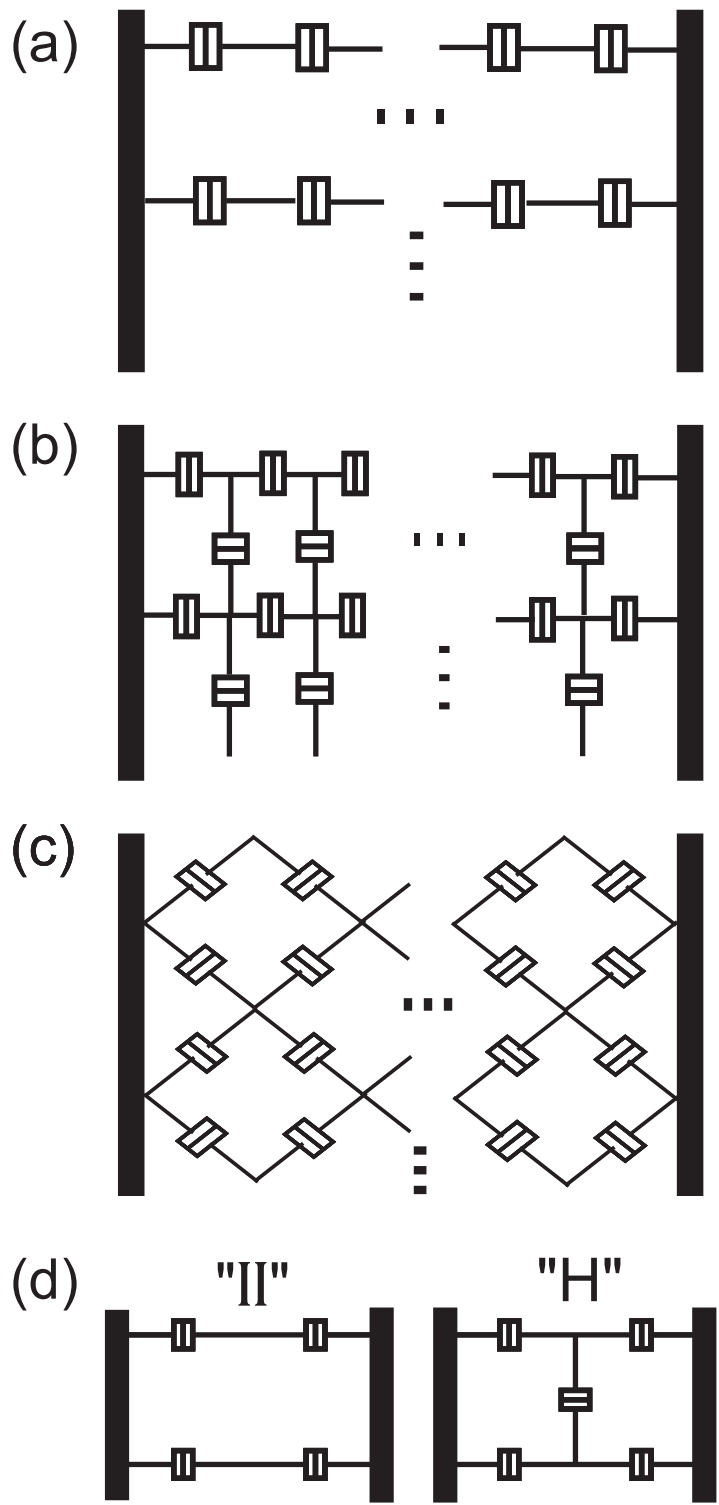
{\bf Fig. 1.} Schematic presentation of the different array structures investigated: {\bf (a)} 1D, {\bf (b)} 
"aligned" 2D, {\bf (c)} "diagonal" 2D arrays, and {\bf (d)} the double junction {\bf II} (left) and the coupled 
double junction {\bf H} (right) structures. 
 
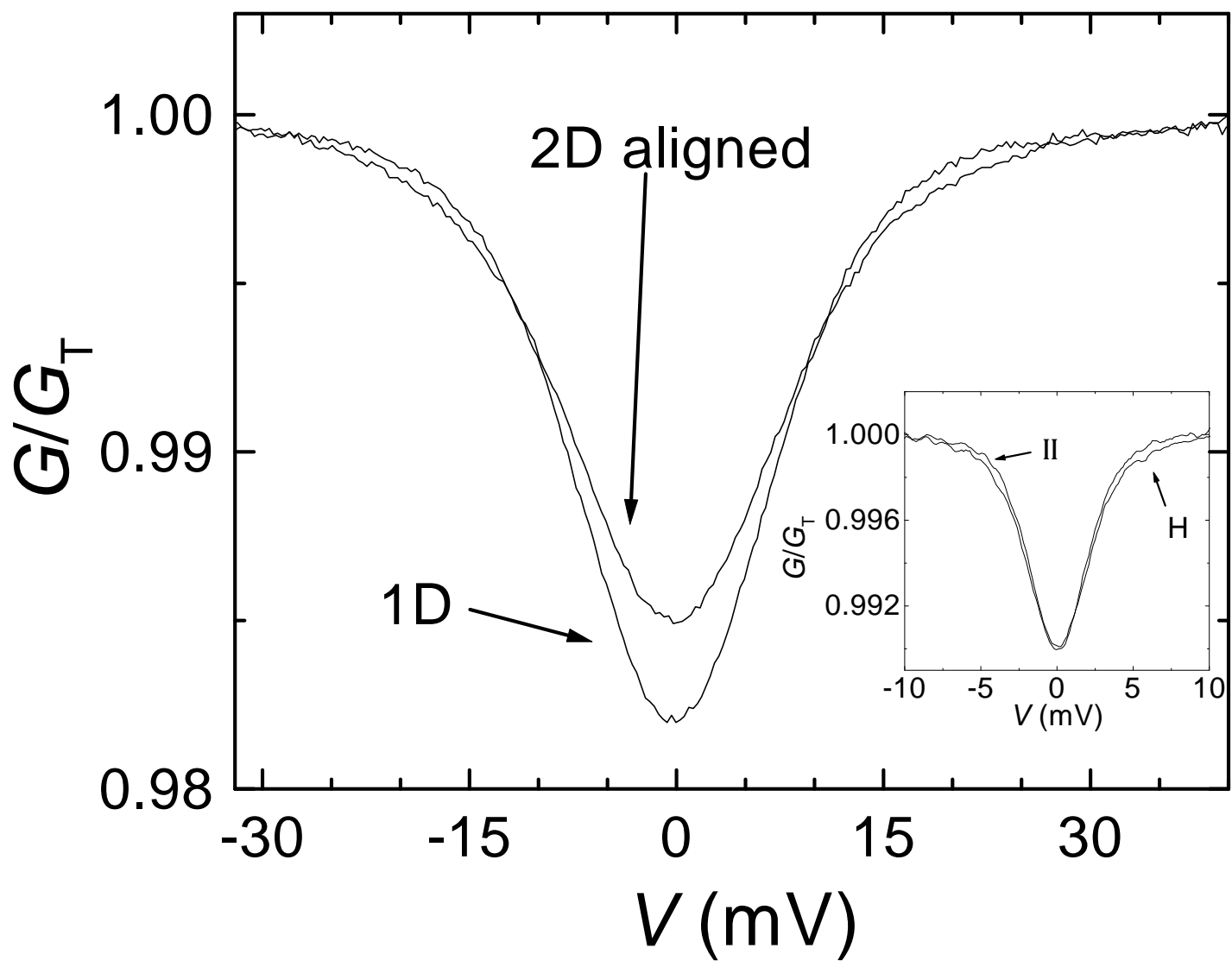
{\bf Fig. 2.} Measured normalised conductance $G/G_T$ vs. bias voltage $V$  for the topologies (a) and (b) of 
Fig. 1 with $N=8$ and $M=9$, measured at 4.25 K. Inset shows the corresponding conductance curves of the 
{\bf II} and {\bf H} structures in Fig. 1 (d). 

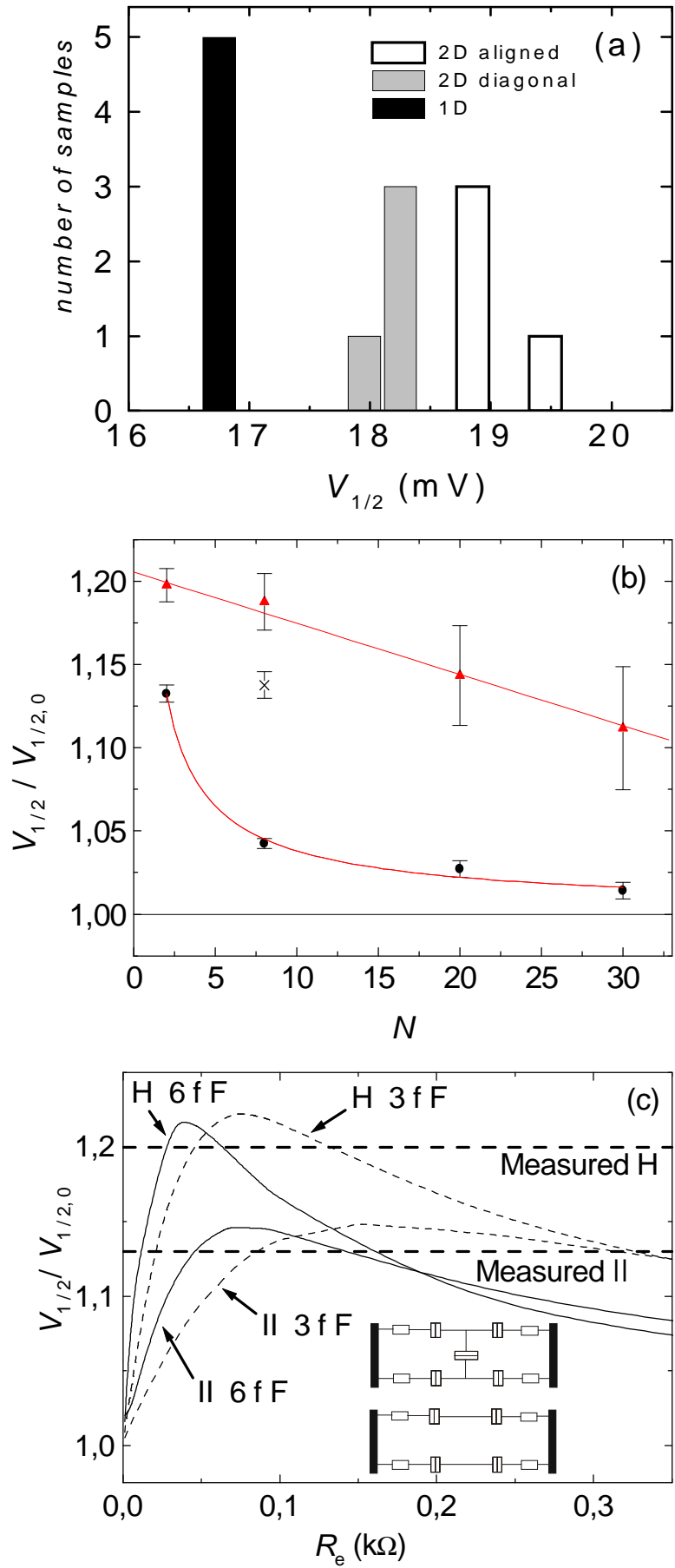
{\bf Fig. 3.} Summary of the experimental and theoretical results: {\bf(a)} Histogram of the measured 
$V_{1/2}$ of the samples with $N=8$, $M=9$, and $M^{\prime}=4$ (at 4.25 K). The filled black bars are for 
1D arrays, the grey ones are for "diagonal" 2D arrays, and the open ones for the "aligned" 2D arrays. {\bf (b)} 
the half width of the conductance curve for 1D (filled circles) and "aligned" 2D (triangles) arrays against $N$. The 
"crossed" 2D data ($N=8$, $M'=4$) are shown by a cross. The lines are just guiding the eye. In {\bf (c)} we 
show the circuit models used in the calculations, and the theoretically obtained half widths for the two structures 
{\bf II} and {\bf H} as functions of the external impedance (resistance) $R_e$. In all cases $T=4.25 $ K, and all 
the junctions are identical with $C=6$ fF ($C=3$ fF) for the solid (dashed) lines. Changing either $C$ or $T$ 
simply scales the values on the horizontal axis; the main parameter in the calculation is $[\hbar/(R_{\rm 
e}C)]/(k_{\rm B}T)$. 

\end{document}